\documentclass{emulateapj}

\newcommand{\etal}{{\it et al.}}
\newcommand{\mf}{M/\Phi_{\rm B}}

\begin{document}
\title{Observational Constraints on the Ages of Molecular Clouds and the
Star-Formation Timescale: Ambipolar-Diffusion--Controlled or
Turbulence-Induced Star Formation?}
\author{Telemachos Ch. Mouschovias, Konstantinos Tassis\altaffilmark{1}, 
and Matthew W. Kunz}
\affil{Departments of Physics and Astronomy,
University of Illinois at Urbana-Champaign, 1002 W. Green Street, Urbana, IL
61801}
\altaffiltext{1}{Current Address: Department of Astronomy and Astrophysics 
and the Kavli Institute for Cosmological Physics, 5640 South Ellis
Avenue, University of Chicago, Chicago, IL 60637}

\begin{abstract}

We revisit the problem of the star formation timescale and
the ages of molecular clouds. The apparent overabundance of
star-forming molecular clouds over clouds without active star
formation has been thought to indicate that molecular clouds are
``short-lived'' and that star formation is ``rapid''. We show that this
statistical argument lacks self-consistency and, even within the rapid
star-formation scenario, implies cloud lifetimes $\approx 10$ Myr.
We discuss additional observational evidence from external galaxies that
indicate lifetimes of molecular clouds and a timescale of star formation
of $\approx 10^7$ yr. These long cloud lifetimes in conjunction with the
rapid ($\approx 1$ Myr) decay of supersonic turbulence present severe
difficulties for the  scenario of turbulence-controlled star formation.
By contrast, we show that all 31 existing observations of objects for
which the linewidth, the size, and the magnetic field strength have
been reliably measured are in excellent {\em quantitative} agreement
with the predictions of the ambipolar-diffusion theory. Within the
ambipolar-diffusion-controlled star formation theory the linewidths may be
attributed to large-scale non-radial cloud oscillations (essentially
standing large-amplitude, long-wavelength Alfv\'{e}n waves), and the
predicted relation between the linewidth, the size, and the magnetic
field is a natural consequence of magnetic support of self-gravitating clouds.

\end{abstract}


\section{Introduction}

The ages of molecular clouds and the timescale of star formation
are currently at the center of an important debate in the field.
However, the debate has a long history. Early on, Giant Molecular
Clouds (GMCs) were believed to be very long-lived ($> 10^8$ yr)
\citep{sc79}. That estimate relied on two arguments.
First, the distribution of CO emission in
galactocentric coordinates lacked a clearly recognizable spiral
pattern, indicating that GMCs are situated in both arm and
interarm regions \citep{sss79}. This implied that GMCs must be
older than the rotational period of the Galaxy ($\simeq 10^8$ yr).
Second, it was estimated that most of the
interstellar hydrogen in the ``molecular ring'' (4-8 kpc) was
molecular rather than atomic or ionic. Therefore the gas must
spend most of its time in molecular form, which in turn implied
GMC lifetimes greater than $10^8$ yr.

Those early arguments for very long lifetimes of GMCs were refuted
by \citet{bls80}. They showed that the molecular-to-atomic
hydrogen gas ratio was overestimated, while the random motions of
the GMCs were neglected in the estimates of the kinematic distance,
thereby leading to an erroneous spatial mapping. Furthermore, they
presented a number of arguments that set the upper limit on the
ages of GMCs at a few $\times 10^7$ yr.

It has recently been suggested that GMCs are short-lived ($\simeq
10^6$ yr), transient objects \citep{Elmegreen00,HBB01}. The idea
of short-lived molecular clouds has been thought to favor a
scenario of turbulence-controlled star formation over the
ambipolar-diffusion--controlled theory for two reasons. First, if
the lifetime of GMCs is smaller than the ambipolar-diffusion
timescale, then ambipolar diffusion does not have enough time to
operate in molecular clouds and thus cannot be relevant to the
star-formation process. Second, short cloud lifetimes help to circumvent
the problem of rapid dissipation of supersonic turbulence and ease the
energy requirements on the source(s) of turbulence, whatever that(those)
might be.

The idea of short-lived molecular clouds is based on observational
estimates of the ages of newborn stars in star forming regions and
on molecular-cloud core statistics. These observations, however,
were shown to be in excellent quantitative agreement with
predictions of the ambipolar-diffusion theory for the timescales
of the {\em observed} phases of star formation (Tassis \&
Mouschovias 2004; hereafter TM04). In this paper, we examine
additional estimators of the ages of molecular clouds and of the
star-formation timescale, and we present further observational
evidence in favor of cloud lifetimes $\simeq 10^7$ yrs that has
received little or no attention until now. We discuss the
implications that cloud ages $\simeq 10^7$ yr have for current
theories of star formation or for ideas on how the star-formation
process is initiated. We also extend the work of Mouschovias \&
Psaltis (1995) and show that all 31 existing observations of
objects (clouds, cores and even masers) for which the linewidth, 
the size, and the magnetic field
strength have been reliably measured are in excellent quantitative
agreement with the predicted relation between those three
quantities, which is a natural and unavoidable consequence of
magnetic support of self-gravitating clouds (Mouschovias 1987a).

\section{Cloud Statistics and Rapid Star Formation}
\label{cloudstat}

The timescale of star formation derived from estimates of the ages
of protostars and the age spreads of stars in clusters was shown
by TM04 to reflect only the late stages of star formation, after
the creation of an opaque hydrostatic core, and ignores a
potentially long earlier phase. However, the existence of an
appreciably long phase before the appearance of a hydrostatic
protostellar core implies that molecular clouds spend a
significant part of their lives without observable embedded
protostars. If that is the case, a significant fraction of
molecular clouds should be observed to contain no protostars. Yet,
most molecular clouds are observed to have embedded protostars. It
is thus claimed that the statistics of molecular clouds with and
without protostars is at odds with the picture presented by TM04
\citep{KBPVSDR}.

First, it cannot be overemphasized that the duration of this
starless phase, which is essentially the time it takes ambipolar
diffusion to form a magnetically (and thermally) supercritical,
dynamically-contracting core (or, fragment) from the mean density
of an initially magnetically subcritical molecular cloud, is not a
universal number -- although it is often used as if it were such.
It depends: (1) on the factor by which the initial central
mass-to-flux ratio of the parent cloud is smaller than its
critical value for collapse \citep{ms76,m91a,fm93,cb01}; and (2)
on the (initial) degree of ionization of the parent cloud
\citep{m79,m87a,m96}. These two quantities are {\em observational
input} to, not predictions of, the theory. The duration of the
subcritical phase of core formation and contraction can be as
short as 1 Myr for mildly subcritical clouds, and as long as 10
Myr for strongly subcritical clouds. The ambipolar-diffusion
theory does {\em not} predict {\em nor} require strongly
magnetically subcritical clouds. Therefore, it does not require
lifetimes of molecular clouds of the order of 10 Myr in order to
be relevant for star formation. For these reasons, the
ambipolar-diffusion theory of star formation does {\em not}
require molecular clouds to spend most of their lifetimes in a
starless state although, if this turns out to be the case, the
theory can definitely accommodate and explain such an observation.

Second, the claim that most observed clouds contain protostars is
based on a very biased and incomplete list of molecular clouds
(those in the solar neighborhood). If molecular-cloud formation is
triggered by a spiral density shock wave, as envisioned by
\citet{msw74}, and as evidenced by the appearance of young, bright
OB stars downstream from the galactic shock \citep{morg70}, then
young molecular clouds without embedded stars should be found
behind the galactic shock, which is traced (in external galaxies)
by dust lanes and a sharp peak of HI emission. As one observes
matter farther away from the shock {\em across} a spiral arm, one
should see older clouds that begin to give birth to stars. This
picture is indeed confirmed by observations of external galaxies
seen face-on (see \S\,\ref{egs} below). In the Milky Way, such
surveys across spiral arms are difficult, if not impossible, to
perform. Instead, surveys of molecular clouds in the Milky Way are
biased toward active regions of star formation.

Third, even if we were to accept the observations in the solar neighborhood
on which this claim is based as representative, they do not support the idea
of ``young'' molecular clouds. If $\tau_{\rm SF}$ is the star-formation
timescale and $\tau_{\rm MC}$ the molecular-cloud lifetime, then the
statistical argument implies that
\begin{equation}\label{ev}
\frac{\tau_{\rm SF}}{\tau_{\rm MC}}=\frac{N_{\rm NS}}{N_{\rm total}}\,,
\end{equation}
where $N_{\rm NS}$ is the number of molecular clouds with {\em N}o {\em
S}tars, and $N_{\rm total}$ is the total number of observed
clouds. (Note: For $t \leqslant \tau_{\rm SF}$, no cloud contains any stars.
If $\tau_{\rm SF} > \tau_{\rm MC}$, the clouds will disperse before they 
form stars; hence, $N_{\rm NS}=N_{\rm total}$. If $\tau_{\rm SF} \approx 
\tau_{\rm MC}$, the clouds will disperse just as they are ready to form 
stars; hence, $N_{\rm NS} \approx N_{\rm total}$. Eq. [1] holds for 
$\tau_{\rm SF} \leqslant \tau_{\rm MC}$.) 
In the ``rapid'' star-formation scenario, $\tau_{\rm
SF}\simeq 1 {\rm \, Myr}$ \citep{vsksbp05}. Then, if $N_{\rm
NS}/N_{\rm total} \simeq 0.1$, as it has been claimed, equation
(\ref{ev}) yields $\tau_{\rm MC} \simeq 10 {\rm \, Myr}$, and the
clouds are {\em not} young. Note that the smaller the $N_{\rm
NS}/N_{\rm total}$ ratio, the greater the estimated lifetime of
molecular clouds, as given by this statistical argument! However,
such a scenario (small $\tau_{\rm SF}$ and large $\tau_{\rm MC}$)
is contradicted by observations of age spreads of young stars in
star-forming regions, which are found to be a few Myr (e.g., Sung
\etal\, 1998; Baume \etal\, 2003). Hence, the
``rapid'' star formation scenario lacks internal consistency.

A recent, independent line of reasoning \citep{GL05}, that relies
on the measured atomic-hydrogen content of molecular clouds to
estimate the clouds' ages, lends support to lifetimes $\gtrsim
10^{7}$ yr.

\section{Evidence for the Star-Formation Timescale in External Galaxies}
\label{egs}

\citet{RobertsWW67} first suggested that galactic shock waves traced by
dust lanes and the sharp HI emission peak may provide the triggering
mechanism for star formation and thus newborn stars should be located
downstream at a distance implied by the timescale for star
formation and the rotational speed of the gas relative to the spiral
pattern. Early on, \citet{RobertsMS67} showed that the
circumferential bands with highest HI distribution in a number of
spiral galaxies lie significantly outside the bands of the optical
arms containing the most prominent newly-born stars and HII regions.

In M51, \citet{mvb72} did a radio continuum survey at 450 pc
linear resolution at the distance of M51. They found the peak of
radio intensity to coincide with the dust lanes, at the inner edge
of spiral arms, but they detected a lag with respect to the position of
the bright young stars. With respect to the galaxy's center, the spiral
arms delineated by the radio emission and the spiral arms delineated by
the HII regions have a difference of $18^\circ$ in position angle,
at a distance $R$ from the galactic center. This corresponds to a
linear separation of
\begin{equation}
L = 2R\sin \frac{18^\circ}{2} = 0.31R\,.
\end{equation}
From this linear separation, they calculated the time it takes the
gas to reach the position of the bright stars after it encounters
the galactic shock, and they found that the star-formation
timescale is
\begin{equation}
\tau_{\rm SF} = \frac{L}{v_{\rm d}} = 6 \times 10^6
\left(\frac{D}{4{\, \rm Mpc}}\right)\left(\frac{100 {\rm \, km \,
s^{-1}}}{v_{\rm d}}\right)\, \, \, {\, \rm yr}\,,
\end{equation}
where $v_{\rm d}$ is the difference between the rotation speeds of the gas
and the spiral pattern. Using the updated value for the
distance of M51, $D=8.4{\rm \, Mpc}$ \citep{fgj97}, equation (3) yields
\begin{equation}
\tau_{\rm SF} = 12.5 {\, \rm Myr}\,.
\end{equation}
This shift between CO and H$\alpha$ peaks was also observed by
\citet{v88} and by \citet{rk90}. More recently, the molecular
content of M51 was studied by \citet{gb93} with the IRAM 30m
telescope. Their resolution was about 560 pc at the distance of
M51 (they adopted a distance of 9.6 Mpc, from Sandage \& Tamman
1975). The spiral arms are prominently traced by CO with the peak
emission being located at the dust lanes, coincident with the
nonthermal radio emission peak, and is separated from the
H$\alpha$ peak by a distance corresponding to the aforementioned
timescale $>10^7$ yr. The density of the molecular gas was
estimated from the intensity ratio of the J=2$-$1 and J=1$-$0 CO
emission to be typical of GMCs.

Similarly in M81, \citet{Rots75} found a phase difference
$10^\circ$ to $15^\circ$ between the blue dips in the optical
spiral arms and the peaks of HI surface density. Using a speed of
20 km s$^{-1}$ kpc$^{-1}$ for the spiral pattern and an average
radius of 5 kpc, he estimated the time lag to be approximately 10
Myr.

\section{Implications for Star Formation Theories}

Given the arguments above and in TM04 against short-lived ($\simeq
1$ Myr) molecular clouds, it is natural to ask where star
formation theory stands in this regard. It has long been known
that molecular clouds exhibit supersonic linewidths \citep{zp74},
which are inextricably linked to how such dense ($n\simeq 10^3$
cm$^{-3}$), cold ($T\simeq 10$ K) objects, whose masses are
typically $10^2$ to $10^4$ greater than the thermal (or Jeans, or
Bonnor-Ebert) critical mass, could be supported against their
self-gravity, or whether they are supported at all. Possible
explanations for the linewidths are radial collapse (or expansion)
\citep{s73,l74,gk74,ss74}, random motions of clumps within clouds
\citep{ze74,mztp74}, supersonic turbulence \citep{l81,lkm82,m83},
and hydromagnetic waves \citep{am75,m75,zj83}. The first two
possibilities have long been ruled out \citep{ze74,m75}, while
there still exists a debate over the latter two. Regardless of
one's opinion, a common task is at hand: how to maintain the
motions responsible for the large linewidths over a cloud's
lifetime ($\simeq 10$ Myr). Any explanation has direct
implications for cloud structure and evolution and, hence, for
theories of star formation.

\subsection{Turbulence-Induced Star Formation}

When Larson (1981) compiled data on linewidths of 54 clouds,
clumps, and cores, and found a relation between the observed
velocity dispersion $\sigma_{\rm v}$ and the size (diameter) $L$
of each object (the so-called ``turbulence law" $\sigma_{\rm
v}\propto L^{0.38}$, which was thought to be the signature of
Kolmogorov turbulence), supersonic turbulence appeared to be a
natural explanation. Subsequent work by \citet{lkm82} and
\citet{m83} also found a power-law relation, albeit with a
significantly greater exponent, $\simeq 0.5$. Due in part to the
influence of Larson's original work, it is widely believed even
today that the characteristic scaling relations at the heart of
theories of turbulence may still provide the most natural
explanation of the linewidth-size relation \citep{mg99}.

It is well known that supersonic turbulence decays very rapidly 
($\lesssim 1$ Myr)
and has very high energy requirements \citep{ms56,f70,f73}.
Moreover, relatively recent numerical simulations
\citep{sog98,ml98,ogs99,pn99,osg01} show that magnetic fields
cannot mediate the decay of such turbulence (often assumed to be
initially superAlfv\'{e}nic, although such an assumption lacks
observational support; see below).

In light of the arguments for long lifetimes of molecular clouds, this
relatively rapid decay poses a serious problem for the turbulence-induced
star formation idea, that assumes that turbulence alone is responsible
for suppressing global cloud collapse while allowing local collapse to
occur in turbulence-produced cores. A suitable driving mechanism that
can replenish the rapidly-decaying turbulence is required.

Both internal and external driving mechanisms have been considered
as means for replenishing the turbulence. Possible candidates for
internal driving are stellar winds, bipolar outflows, and HII
regions. However, there exist molecular clouds devoid of visible
star formation that exhibit a higher level of turbulence than
clouds with embedded OB associations \citep{mckee99}. Besides, a
theory of star formation must be able to explain how the first
stars form in a molecular cloud, without having to rely on
possible {\em\ stellar} triggers. These arguments, along with
results from simulations of driven cloud turbulence
\citep{om02,KBPVSDR}, have led to the realization that internal
driving cannot be reconciled with observations of actual molecular
clouds. As a result, attention is now turning to external driving
mechanisms, which include supernovae, density waves, differential
rotation of galactic disks, and winds from massive stars.

External driving of supersonic turbulence has its own
serious difficulties:

1. Inward propagating disturbances (compressible turbulence) 
impart linear
momentum to the matter in their direction of propagation, thereby
tending to aid the self-gravity of a cloud in inducing collapse
\citep{m87a}. Consequently, a mechanism other than turbulence must
exist and be responsible for the support of self-gravitating
clouds. 

2. The material motions (e.g., shocks) implied by the inward
propagating, nonlinear disturbances ($\simeq 1$ ${\rm km \, s^{-1}}$)
even in clouds that have not yet given birth to stars are not observed 
in molecular clouds.

3. External driving from supernovae (or the winds of massive
stars), which are currently considered to be the most promising
external source of turbulent energy \citep{KBPVSDR}, assumes that
stars in the neighborhood of a cloud under study have formed by
means unrelated to turbulence. If star formation can take place in
other clouds without any assistance from turbulence, there is no
reason to postulate that the same mechanism responsible for that
star formation cannot operate in the cloud under consideration. 

Regardless of the possible sources of supersonic turbulence or any
role that such turbulence may or may not play in the structure and evolution
of molecular clouds, the linewidth-size relation that was a key
motivation for invoking such turbulence in the first place has not
been adequately explained by numerical simulations of turbulence
(see \S\,5.2 in Elmegreen \& Scalo 2004 for a discussion of the wide
range of conflicting results on linewidth-size relations from different
numerical studies).

\subsection{Ambipolar-Diffusion--Initiated Star Formation}

\subsubsection{Theory vs. Observations}

The role of cosmic magnetic fields in the formation and support of
self-gravitating clouds and the formation and evolution of
protostars in such clouds has been synthesized into the theory of
ambipolar-diffusion--initiated star formation, based on detailed
analytical and numerical work over the last thirty years (see
reviews by Mouschovias 1978, 1981, 1987a, 1987b, 1995,
1996; Mouschovias \& Ciolek 1999). In this theory, the formation
of self-gravitating cloud cores (or fragments) and their evolution
into protostars is a result of the incessant struggle between
gravitational forces and their principal opponent, magnetic
forces, with ambipolar diffusion being the clever means by which
gravity eventually wins that battle.

The outcome of this struggle is determined mainly by the initial
mass-to-flux ratio, $\mf$, and the initial degree of ionization,
$x_{\rm i}\equiv n_{\rm i}/n_{\rm n}$, of the parent {\em cloud}.
These quantities are not predictions of the theory; they are {\em
observational input} to the theory. (However, the theory makes
definite predictions about the mass-to-flux ratio of molecular
cloud {\em cores}; namely, it should typically be supercritical by
a factor 1 - 3 for cores with central density $\simeq 10^{5} -
10^{9} \, {\rm cm^{-3}}$; e.g., see Fiedler \& Mouschovias 1993,
Fig. 9b. This prediction is in excellent agreement with all
observations of mass-to-flux ratios in cloud cores to date
[Crutcher 1999; Crutcher {\em et al.} 2004; Heiles \& Crutcher 2005].
For the effect of rotation and/or grains on this prediction, see,
respectively, Basu \& Mouschovias 1994, Fig. 8b; Ciolek \&
Mouschovias 1994, Fig. 4e.)
Ambipolar diffusion leads to {\em quasistatic} (i.e., negligible
acceleration, but not necessarily negligible velocity) formation
of magnetically supercritical cores in the deep interiors of
molecular clouds followed by their {\em dynamic} contraction
(collapse, but not free fall). The envelope of the parent
molecular cloud remains magnetically supported. The theory
predicts ordered large-scale magnetic fields with hourglass
morphology and with strength $B$ in the cores that is related to
their density $\rho$ by $B\propto \rho^{\kappa}$, where
$\kappa=0.47$. Thermal pressure forces are primarily responsible
for maintaining a nearly uniform density in the central region,
while the density decreases approximately as $\rho\propto r^{-2}$
in a region $r \approx 10^2-10^4$ AU, the dynamically contracting
(but not free-falling), thermally and magnetically supercritical
core. (This structure is distinct from Shu's [1977]
singular isothermal sphere model of low-mass [a few solar masses]
clouds, in which the $r^{-2}$ density profile refers to the {\em
static} cloud envelope, the dynamically contracting inner region
has a $r^{-1.5}$ profile, and there is no flat-density central
region.) Turbulence is not required for cloud support (although,
if present, it is accommodated in the theory), so the problem of
its rapid decay is irrelevant to the ambipolar-diffusion theory of
star formation.

Observations of molecular clouds confirm many of the predictions
of the ambipolar-diffusion theory and, just as importantly,
contradict none. The envelopes of molecular clouds are subcritical
(or critical), whereas cloud cores are supercritical
\citep{ctghkm93,cmtc94,crmt96,cnwtk04,hc05,ccw05}. Large-scale
ordered fields, often with hourglass morphology, are seen
threading the clouds \citep{vct81,s98,lcgr02} and their cores.
Low-mass cores are found in the deep interiors of clouds, rather
than near the surface \citep{jdfk04}. The observed scaling
relation between $B$ and $\rho$ has $\kappa=0.47 \pm 0.08$ \citep{c99}.
Millimeter and submillimeter continuum observations by
\citet{wtsha94} and \citet{awtm96} yield density profiles for
starless cores that are in excellent agreement with those
predicted by the ambipolar-diffusion theory. Ion-neutral drift
speeds have been measured by \citet{bcm98} and found consistent
with the theory. Specific dynamical models have been constructed
for the Barnard B1 cloud \citep{cmtc94} and L1544 \citep{cb00a}
and have predicted core properties that are in excellent agreement
with observations. \citet{cb00b} discuss further the close
agreement between the ambipolar-diffusion theory and observations.
By contrast, {\em quantitative} agreement between observations and
the results of simulations resulting in presumed initiation of
fragmentation (or core formation) by turbulence in molecular
clouds is lacking.

Two points must be made regarding the mass-to-flux ratio. First,
statistical arguments have been offered (e.g., Crutcher 1999) to the effect
that clouds are supercritical, in presumed contradiction to the ambipolar
diffusion theory. However, all observations on which that deduction is
made are based on measurements of the magnetic field strength in {\em cores}.
As explained above, the definite prediction of the
ambipolar-diffusion theory has been that typical cores {\em must} 
be nearly critical or 
slightly supercritical. Hence, the criticism of the
ambipolar-diffusion theory on the basis of observed mass-to-flux
ratios rests on a fundamental misunderstanding of the theory.
\footnote{This confusion has found its way into recent papers
[e.g., Mac Low \& Klessen (2004), Li \etal\, (2004)] that criticize
the ambipolar-diffusion theory on the grounds that (1) magnetic
clouds {\em as a whole} are observed to be supercritical, and (2)
the ambipolar-diffusion theory predicts subcritical cores, {\em
neither of which is true}. Similar misinterpretation of
observational data has also led to simulations of supercritical
molecular clouds, sometimes with mass-to-flux ratios an order of
magnitude supercritical \citep{lnmlh04}. Such a large mass-to-flux
ratio has never been observed in molecular clouds.}

Second, there is an important distinction between the
ambipolar-diffusion theory of star formation and the idea of
turbulence-driven star formation as they relate to the
mass-to-flux ratio. The theory of ambipolar-diffusion--initiated
star formation predicts tight constraints on how the mass-to-flux
ratio should vary between a cloud envelope and its cores (see
above). Turbulence-driven fragmentation, on the other hand,
predicts $\mf\gg (\mf)_{\rm crit}$ to be as equally likely as
$\mf\approx (\mf)_{\rm crit}$ in {\em both} molecular cloud cores
{\em and} their envelopes. Observations show that this is not the
case (e.g., see Heiles \& Crutcher 2005).

\subsubsection{Explanation of the Linewidths}

Magnetic support of molecular clouds has led to a natural explanation
of the observed linewidths, without any {\em ad hoc} assumptions.
If (1) self-gravitating clouds are magnetically supported, and (2) the
material velocities responsible for the supersonic linewidths are
slightly sub-Alfv\'{e}nic or Alfv\'{e}nic [i.e., $(\Delta v)_{\rm
NT}\simeq(\Delta v)_{\rm wave}\lesssim v_{\rm A}$], then the
linewidths may be attributed to large-scale non-radial cloud
oscillations, which are essentially {\em standing}
large-amplitude, long-wavelength ($\lambda \simeq 1$ pc)
Alfv\'{e}n waves. The predicted nonthermal linewidth ($\Delta v)_{\rm NT}$
is related to the magnetic field strength $B$ and the size $R$ of the
object by
\begin{equation}\label{linewidth}
(\Delta v)_{\rm NT} \simeq 1.4\left(\frac{B}{30\;\mu{\rm
G}}\right)^{1/2}\left(\frac{R}{1\;{\rm pc}}\right)^{1/2}\; \, \, \, {\rm
km\; s}^{-1}
\end{equation}
(Mouschovias 1987a; Mouschovias \& Psaltis 1995). Equilibrium oscillations
left over from the cloud formation
process \citep{m75,g05} could be the origin of these waves. They decay
on the ambipolar-diffusion timescale.

That the linewidths should be slightly subAlfv\'{e}nic or Alfv\'{e}nic
is suggested by the work of \citet{mm85a,mm85b}, who found that
equipartition is established between the magnetic energy in
Alfv\'{e}n waves and the kinetic energy of the material motions
associated with the waves after only a few reflections off
fragments (or the cloud surface) and the consequent wave-wave
interaction. Observations of linewidths confirmed that prediction
\citep{mg88,c99}. SuperAlfv\'{e}nic linewidths have {\em never}
been observed in self-gravitating molecular clouds;
this has particularly devastating
consequences for star formation ideas that require
superAlfv\'{e}nic turbulence in order for simulations to
match observed characteristics of clouds (e.g., Padoan \& Nordlund
1999; Li \etal\, 2004).

Equation (\ref{linewidth}) was shown by \citet{mp95} to be in
excellent quantitative agreement with spectral line observations
of clouds, cores, and embedded OH masers; they considered the 14
objects for which the linewidth, size, and magnetic field strength
were measured at the time. By now there are 31 sources for which
all three quantities [($\Delta v)_{\rm NT}$, size $R$, and field
strength $B$] have been reliably measured
\citep{mg88,c99,cnwtk04}. The total magnetic field strength $B$
is taken to be the usual statistical average over inclination
angles: $B=2|B_{\rm los}|=4|B_{\rm sky}|/\pi$ (where $B_{\rm
los}$ and $B_{\rm sky}$ are the components of the field along the
line-of-sight and in the plane of the sky, respectively) for
clouds and cores, and $B=|B_{\rm los}|$ for masers.

We use this data to plot $(\Delta v)_{\rm NT}$ versus $R$ in
Figure 1a. Error bars are as in \citet{mg88}; they indicate an uncertainty
of a factor of 2. No single power law can meaningfully fit the
data, in that the standard deviation would be too large. 
In Figure 1b we separate these points into
weak-field ($B \le 270\;\mu$G; open circles), moderate-field
($270\;\mu$G $<B< 3000\;\mu$G; grey circles), and strong-field
($3000\;\mu$G $\le B$; black circles) regimes. There is a clear
indication that sources of different magnetic-field strength
follow different scaling laws. In Figure 2, we show the same data,
but we plot the quantity $(\Delta v)_{\rm NT}/R^{1/2}$ against
$B$. Error bars are as in Figure 1. The solid line is the theoretical
prediction, equation (\ref{linewidth}). The dashed line is a
least-squares fit to the
data. The quantitative agreement between theory and observations is
remarkable. The theoretical prediction and the least-squares fit
have exactly the same slope. In addition, the fact that the
theoretical prediction is offset slightly higher than the
least-squares fit indicates that the material motions responsible 
for the linewidths are slightly subAlfv\'{e}nic.

\begin{figure}
\centering
\plotone{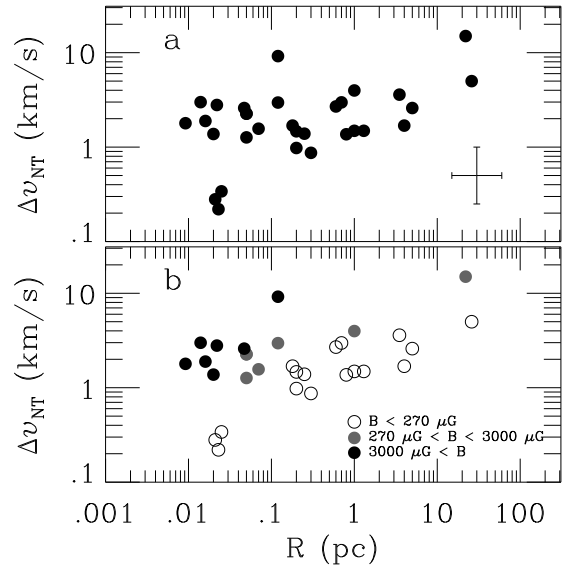}
\caption{(a) Nonthermal linewidth vs. (FWHM) size for 31 objects
(data from Myers \& Goodman 1988; Crutcher 1999; and Crutcher
{\em et al.} 2004). (b) Same linewidth-size data as in (a), but grouped
according to the total magnetic field strength (see text for
definition of $B$).}
\end{figure}

\begin{figure}
\centering
\plotone{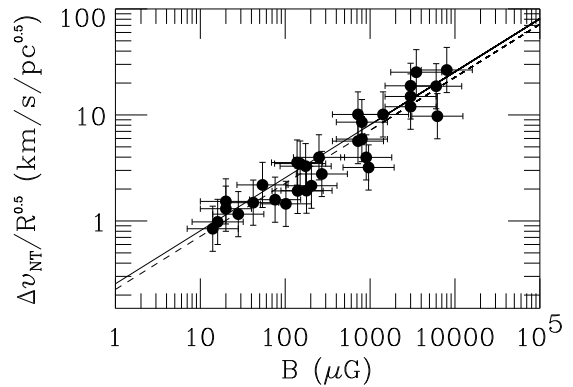}
\caption{Same linewidth-size data as in Fig. 1, but exhibiting the
ratio $(\Delta v)_{\rm NT}/R^{1/2}$ as a function of the total
magnetic field strength $B$. Error bars are as in Fig. 1. The
theoretical prediction (eq. [\ref{linewidth}]) is shown as a solid
line. The dashed line is a least-squares fit to the data.}
\end{figure}

Attributing the linewidths to standing, large-amplitude,
long-wavelength Alfv\'{e}n waves does not affect the evolution of
a protostellar core in any way. For typical molecular cloud
parameters, the size of the region that can just become
gravitationally unstable because of ambipolar diffusion happens to
be essentially equal to the Alfv\'{e}n lengthscale $\lambda_{\rm
A}$ (Alfv\'{e}n waves with wavelengths $\lambda\le\lambda_{\rm A}$
cannot propagate in the neutrals because of damping by ambipolar
diffusion --- see Mouschovias 1991b, eqs. [18a,b]). In fact, it is
precisely the decay of hydromagnetic waves due to ambipolar
diffusion that removes part of the support against gravity over
the critical thermal lengthscale and thus initiates fragmentation
(or core formation) in molecular clouds \citep{m87a}. One should
therefore expect thermalization of linewidths on the Alfv\'{e}n
lengthscale in supercritical cores. This has been observed by
\citet{bapabwt02} in L1544.

\section{Discussion}

In this paper, we have re-examined the argument that an observed
overabundance of molecular clouds that are actively forming stars
with respect to clouds without active star formation indicates
that molecular clouds are ``short-lived'' and that star formation is
``rapid''. According to that argument, ambipolar diffusion does
not have enough time to operate in molecular clouds. We have shown
that: (a) Even if the observational facts used to support that
argument were unbiased and accurate, these statistics imply
molecular cloud lifetimes $\simeq 10$ Myr, even within the ``rapid''
star-formation scenario. (b) Observations of molecular clouds in
the solar neighborhood are {\em not} in fact unbiased or
representative --- rather, quiescent clouds are expected to be
found mostly close to a galactic shock along spiral arms, as is
observed in external, face-on spiral galaxies. (c) The
ambipolar-diffusion theory of star formation does {\em not} require
long quiescent periods of molecular clouds or cloud lifetimes
$\gtrsim 10$ Myr, although it can certainly explain and
accommodate such a possibility. Furthermore, if molecular clouds
were short-lived, ``transient'', ``evanescent'', nongravitating
structures \citep{Elmegreen00,HBB01}, then they should be dispersing
fast, in $\simeq 1$ Myr. This implies that molecular clouds with
embedded protostars should exhibit large, (ordered) expansion
velocities. Such velocities are {\em not} observed.
\footnote{Even if it {\it were} the case that all molecular clouds had 
embedded stars, the conclusions one would draw within the 
ambipolar-diffusion theory would be: (1) The cloud ages are longer than 
the star-formation timescale. (2) The ambipolar-diffusion timescale in 
every molecular cloud is smaller than the cloud's lifetime. (3) Molecular 
clouds formed at about the same time (behind a spiral density shock wave). 
Consequently, statistical arguments cannot be applied to such clouds.}

We have also examined the implications of these results for
star-formation theories or ideas. Theories or ideas that depend on
supersonic, magnetized or non-magnetized turbulence for
fragmentation, core formation, and cloud support are burdened by
the requirement of continuously replenishing the turbulent motions during the
entire lifetime of a molecular cloud.
Proponents of this idea have found that internal driving of the
turbulence cannot be reconciled with observations of actual
molecular clouds. We have argued that external driving will most
likely result in cloud compression and premature collapse. Also,
the most promising external driving mechanisms (e.g., supernovae
and winds from massive stars) assume the pre-existence of
stars, and do not explain the origin of that previous generation
of stars. Furthermore, recent observations \citep{kshs05},
which find molecular clouds to be preferentially elongated
along the Galactic plane, are in conflict with driving by supernovae
and stellar winds, since these cannot account for the preferred elongation.
Without an adequate source of driving, the observed supersonic
linewidths cannot be maintained over a cloud's lifetime. Moreover,
the linewidth-size relation first established by \citet{l81} and
extended by \citet{lkm82} and \citet{m83}, which was a key
motivation for invoking turbulence in the first place, has not
been explained by current numerical simulations of molecular cloud
turbulence. Since superAlfv\'{e}nic linewidths have never been
observed in self-gravitating molecular clouds, turbulent star
formation ideas that require superAlfv\'{e}nic turbulence in order
for their simulations to match observed cloud properties (e.g,
Padoan \& Nordlund 1999; Li \etal\, 2004) have no relevance to
actual molecular clouds.

By contrast, the theory of ambipolar-diffusion--initiated star
formation has, over the last thirty years, made numerous {\em
quantitative} predictions that turned out to be in excellent 
agreement with observations.
Cloud envelopes are supported by magnetic fields while ambipolar
diffusion allows supercritical cores to form and dynamically
collapse in the deep interiors of self-gravitating clouds. This is
supported by detailed numerical simulations as well as by
polarimetry and Zeeman observations. The linewidths are due to
nonradial cloud oscillations, which are essentially standing
large-amplitude, long-wavelength Alfv\'{e}n waves. Linewidth,
size, and magnetic field data from 31 clouds, cores, and embedded
masers are in excellent quantitative agreement with this theory.
The theory does not suffer from a need to replenish turbulent
motions in order to support clouds against collapse or to explain
the linewidths, although if such replenishment takes place it has
no effect on the ambipolar-diffusion theory of fragmentation and
star formation: ambipolar diffusion damps the waves precisely over
the lengthscales necessary for gravitational formation and
contraction of fragments (or cores) (see Mouschovias 1987a, 1991).

We caution against using mass-to-flux ratios from observations of
molecular cloud cores to make statements about the mass-to-flux
ratios of molecular cloud envelopes. Supercritical cloud cores are
a prediction of the ambipolar-diffusion--initiated star formation
theory, and do not imply that magnetic support is insignificant
in the envelopes. Moreover, the fact that observed
cloud cores are critical or slightly supercritical and cloud
envelopes are subcritical contradicts the simulations of
turbulence-induced star formation, whose results imply that 
highly supercritical cores are equally likely
as slightly supercritical cores.

It is important to clarify the semantic difference between ``rapid''
and ``slow'' star formation. When one refers to a process as being rapid
or slow, one must also specify with respect to what. The main
theoretical motivation for pursuing the idea of ``short-lived''
molecular clouds was to test whether ambipolar diffusion has
enough time to operate and form supercritical cores over the
lifetime of a molecular cloud. We have once again pointed out that
the timescale of the ambipolar-diffusion--controlled core formation
process depends on the mass-to-flux ratio and the degree of ionization 
of the parent cloud and it can be as short as 1 Myr for
mildly subcritical clouds. Hence, from this perspective, clouds
with ages of a few Myr are not ``young'' enough to render the
ambipolar diffusion theory irrelevant to star formation. Loose terminology
should not replace quantitative standards for determining the validity of
a theory.

{\bf Acknowledgments}: This work has been carried out without
external support, and this paper would not have been published without the
generosity of {\em The Astrophysical Journal}.


\end{document}